\documentclass[rspublic,showpacs,twocolumn,superscriptaddress,groupedaddress]{revtex4-1}
\usepackage{bm}        
\usepackage{amssymb}   
\usepackage{graphicx}
\usepackage{epstopdf}
\usepackage{color}
\usepackage{amsmath,amssymb}
\usepackage{footmisc}

\begin{document}

\title{Increased Network Interdependency Leads to Aging}



\author{Dervis Can Vural}
\author{Greg Morrison}
\author{L. Mahadevan}
\affiliation{Department of Physics, School of Engineering and Applied Sciences, Harvard University, Cambridge, MA 02138, USA}


\date{\today}

\begin{abstract}
Although species longevity is subject to a diverse range of selective forces, the mortality curves of a wide variety of organisms are rather similar. We argue that aging and its universal characteristics may have evolved by means of a gradual increase in the systemic interdependence between a large collection of biochemical or mechanical components. Modeling the organism as a dependency network which we create using a constructive evolutionary process, we age it by allowing nodes to be broken or repaired according to a probabilistic algorithm that accounts for random failures/repairs and dependencies. Our simulations show that the network slowly accumulates damage and then catastrophically collapses. We use our simulations to fit experimental data for the time dependent mortality rates of a variety of multicellular organisms and even complex machines such as automobiles. Our study suggests that aging is an emergent finite-size effect in networks with dynamical dependencies and that the qualitative and quantitative features of aging are \emph{not} sensitively dependent on the details of system structure.
\end{abstract}

\maketitle 

\section{Introduction}

For a collection of $s$ radioactive atoms, the probability of decay is a constant, so that the fraction of atoms that decays per unit time $-(ds/dt)/s = \mu$ does not change in time. In other words, an ``old'' atom is equally likely to decay as a ``young'' one. Contrastingly, in complex structures such as organizations, organisms and machines, one finds that the relative fraction that dies per unit time, the mortality or aging parameter $\mu(t)$ varies, and typically increases in time. In living systems $\mu(t)$ increases exponentially (commonly known as the Gompertz Law) up until a ``late-life plateau'', after which aging decelerates. Moreover, the functional form of $\mu(t)$ for a wide variety of organisms is remarkably similar \cite{horiuchi, caughley, vaupel}. 

The origins of biological aging has been sought in two broad classes of theories \cite{hughes,weinert}, which may be non-exclusive. The first, mechanistic approach, aims to understand aging in terms of mechanical and biochemical processes such as telomere shortening \cite{blackburn} or reactive oxygen species damage \cite{harman}. The second approach, considers aging as the outcome of evolutionary forces \cite{baudisch}. The early evolutionary theories are based on the observation that selective pressure is larger for traits that appear earlier in life \cite{medawar2, williams, hamilton}. As a result, aging, it has been argued, could be due to late-acting deleterious mutations accumulated over generations (mutation accumulation theory, MA) \cite{medawar2, medawar1} or due to mutations that increase fitness early in life at the cost of decreasing fitness later in life (antagonistic pleiotropy theory, AP) \cite{williams, rose91, kirkwood02}. Physiological variants of AP consider the relative energy cost of avoiding aging damage versus reproducing \cite{kirkwood77, abrams}: Mutations diverting energy away from repair and maintenance activities to earlier sexual development and high reproduction rate can be favored. 
\begin{figure}
\centering
\includegraphics[width=3.4in]{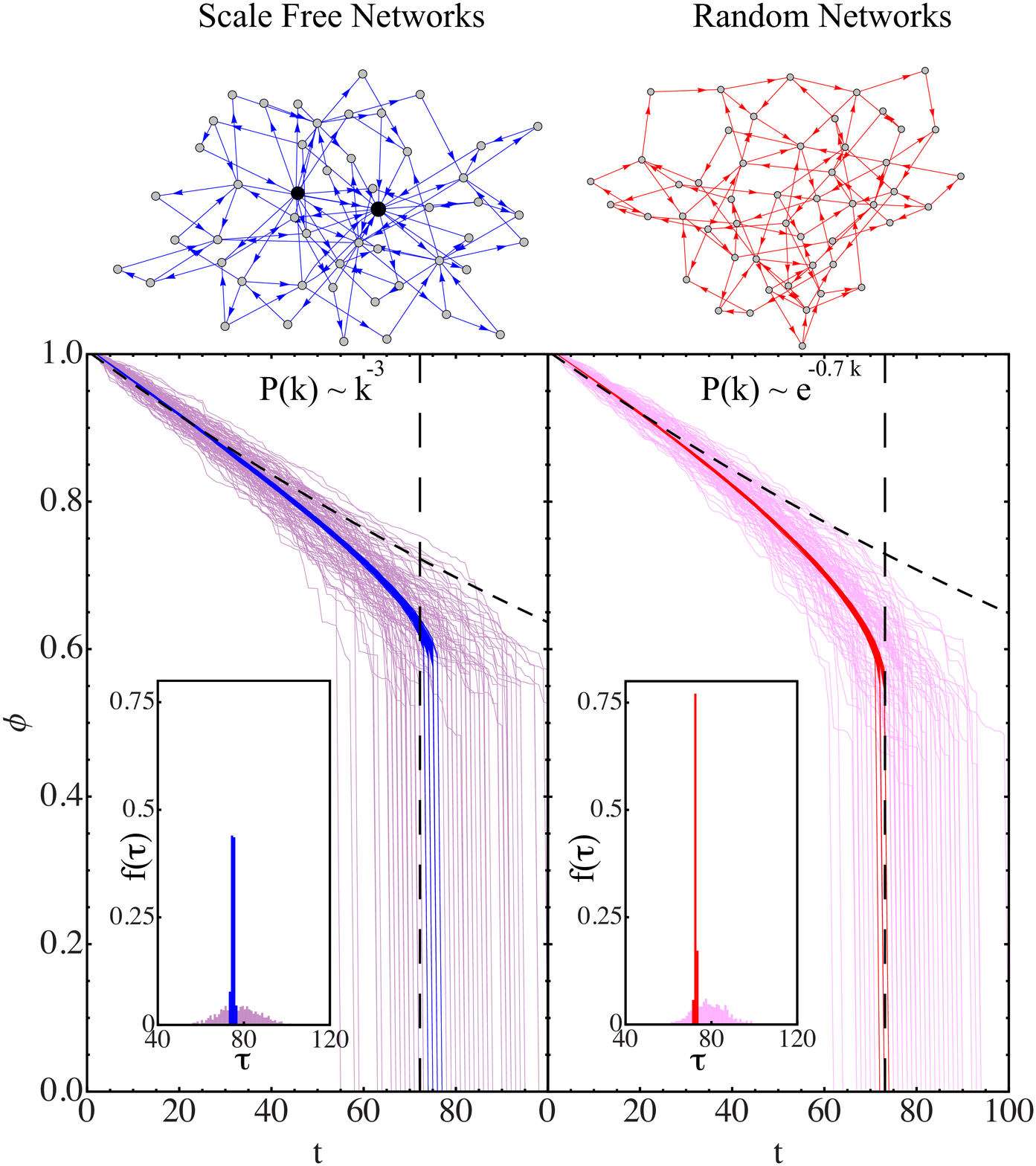}
\caption{\small{\bf Fraction $\phi(t)$ of functional components at time $t$ as a function of dependency network size and topology, 100 Runs}. Dependency structures grown via non-neutral (left column) and neutral (right column) evolutionary schemes yield scale free $P(k)\sim 1/k^3$ and random $P(k)\sim e^{-k/\kappa}$ degree distributions respectively. A representative network for each scheme is shown on top of the respective columns; preferential attachment produces high-degree hubs (large, black nodes) that are absent using random attachment. We plot 100 runs of $\phi(t)$ for each network size and topology and show their respective lifetime $\tau$ distributions $f(\tau)$ in the inset. Increasing the network size from $N=2500$ (purple and pink) to $N=10^6$ (blue and red) sharpens $f(\tau)$. The dashed black lines mark analytical predictions (cf. theory section) for initial slope $p_0$ (which is 1.80 for SFN and 1.75 for RN) and critical fraction $\phi_c$ (which is 0.6 for SFN and 0.5 for RN) which agree well with simulation results. For both plots \{$\gamma_0,\gamma_1,d\}=\{0.0025, 0, 0\}$. Note the remarkable similarity of different network topologies.
}
\end{figure}

A more recent collection of evolutionary theories view aging as a group selection effect (GS); e.g. it has been argued that aging might enhance the rate of evolution by decreasing generation overturn times \cite{libertini}, reduce epidemic outbreaks by diluting the population \cite{mitteldorf} or decrease kin competition between parent and progeny \cite{acrmartin}. 

While the mechanistic theories are firmly supported by experiment and continue to provide insight, the evolutionary origins of aging remain a mystery. The neutral (i.e. non-selective, non-directional) MA theory has two predictions: a monotonic increase in the mortality rate with age, and an increased variation (spread) in mortality rate among different polymorphisms with age, both of which disagrees with observation \cite{curtsinger,carey,promislow}. The non-neutral (i.e. selective, directional) AP theory predicts that every aging gene comes with an early-life enhancement of fecundity. While some such genes have been found, others contradict this prediction \cite{dgray, miller1, miller2, reznick}. 

In contrast, the ideas of GS remain untested. Even if true, these theories do not explain why the mortality rate should increase in time, let alone why so many different organisms should exhibit a similar functional form for $\mu(t)$.


Here we approach the question of aging from an evolutionary perspective, combining ideas from network connectivity \cite{barabasi,stanley1,cohen, stanley2}, engineering reliability analysis as applied to aging \cite{gavrilov,laird}, and the theory of constructive neutral evolution \cite{gray,lynch,stoltzfus}. Much like the historical development of an engineered technological device, the complexity of life appears to have irreversibly increased as a large number of individual sub-units become linked through specialized interdependence. Our simulations and analysis allows us to demonstrate that component dependencies established over the course of neutral and non-neutral evolution eventually leads to species with aging curves that are consistent with empirical data.

\section{Algorithm for network evolution}
To study the dynamics of aging quantitatively, we start with a simple view of an organism as a set of nodes with dependencies characterized by directed edges between them. Each node may be thought as genes in a regulatory network, or the differentiated cells or tissues in a multicellular organism with specific functions. A directed edge from node $A$ to node $B$ indicates that $A$ provides something to $B$ such as energy, crucial enzymes or mechanical support, so that the function of $B$ relies on the function of $A$. In this scenario, the evolution of the network is represented by random addition of new nodes and edges, leading to a change in the dependency structure of the network. This then changes the susceptibility of the network to further changes, including its longevity.
 
\begin{figure}
\centering
\includegraphics[width=3.2in]{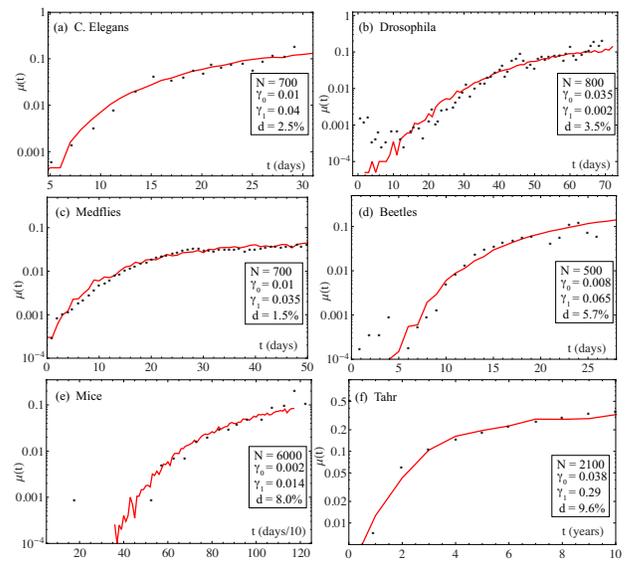}
\caption{{\bf Empirical mortality data (black) fit to present theory (red)} The mortality curves of (left to right) \emph{C. elegans, Drosophila}, Medflies, Beetles, Mice, Himalayan Goats from \cite{horiuchi, caughley, vaupel}. Fit parameters are {\tiny $\{N,\gamma_0,\gamma_1,d\}=$, $\{700, 0.01, 0.04, 0.025\}$, $\{800, 0.0035, 0.002, 0.035\}$, $\{700, 0.01, 0.035, 0.015\}$, $\{500, 0.008, 0.065, 0.057\}$, $\{6000, 0.002, 0.014, 0.08\}$, $\{2100, 0.038, 0.29, 0.096\}$}. The horizontal axis (as well as the units of $1/\gamma_{0,1}$) is time in units of days for \emph{C. elegans}, \emph{Drosophila}, medflies and beetles; days/10 for mice and years for Tahr. To demonstrate that aging is a manifestation of component interdependence we also fit our model to two kinds of cars (cf. supplemental information and Fig.S.1)}
\end{figure}
For a given network, we assume that its dynamics are governed by four parameters, of which three are subject to direct experimental control: The failure rate $\gamma_0\ll1$ and repair rate $\gamma_1\ll1$ of individual nodes, the initial fraction of damaged nodes $d\ll1$, and total number of nodes $N\gg1$. $\gamma_0$ is controlled by biomolecular processes subject of mechanical theories of aging (e.g. oxidative stress, radiation damage); the damage due to pre or post natal stress is contained in $d$. $\gamma_1$ depends on the activity or inactivity of genes or their regulators that may be relevant for repair and replacement of cells. It seems difficult to determine or modulate $N$, though it should roughly correlate with the ``complexity'' of an organism.

The structure of networks is expected to influence the dynamics of any processes on it, including aging. However we surprisingly observe that different evolutionary processes for network growth lead to similar average lifespans and mortality curves. At either extreme of network structure, we use two strategies to grow them via what may be termed neutral and non-neutral evolution. In either case, a new node is introduced at every ``evolutionary time step'' such that it depends on one random existing node, and one random existing node depends on it. In the \emph{constructive neutral} process, the new node is equally likely to depend on and provide to any of the existing nodes. This yields a random dependency network (RN). In the \emph{constructive non-neutral} process  the probability that a new component will provide or receive from another component with degree $k$ is taken proportional to $k$. This yields a scale-free dependency network (SFN) \cite{barabasi1}. In the first scheme there is no difference in fitness  based on whom a node relies on; whereas in the second scheme, relying on a previously unrelied component has a larger fitness cost, making such connections unlikely.  Each network topology is diagrammed schematically in Fig. [1].  

\begin{figure*}
\centering
\includegraphics[width=5.6in]{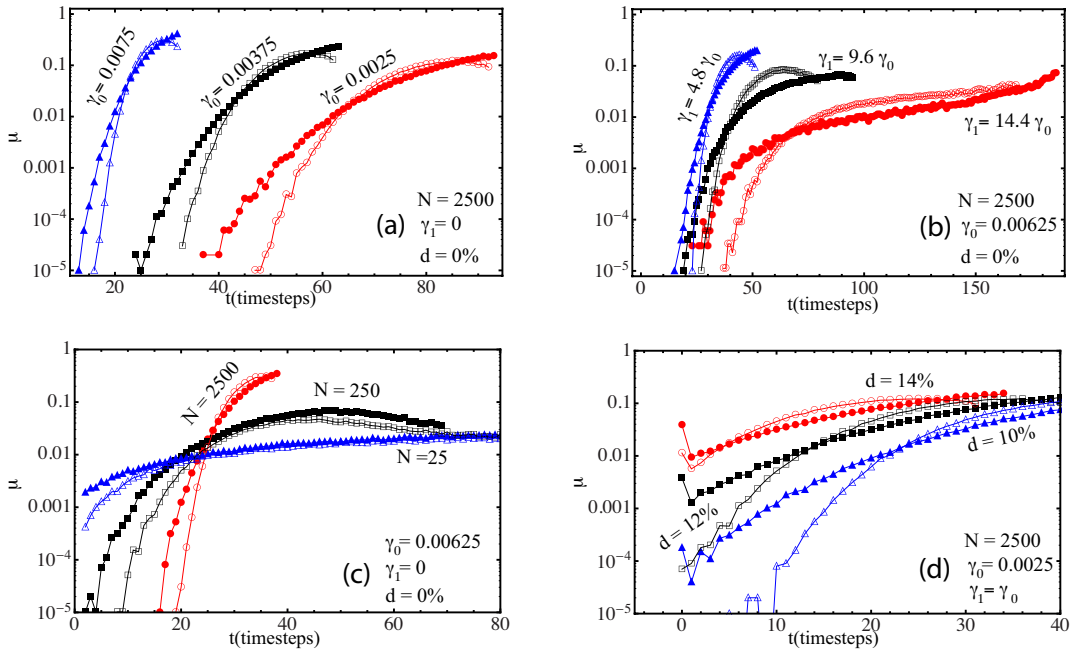}
\caption{{Mortality Rate $\mu(t)=-\partial_ts/s$ as a function of time.} In each panel we test the effect of one single parameter, keeping the others constant. While random networks (open markers) seem to age slightly faster compared to a scale free networks (filled markers), network topology does not seem to have a significant effect on the qualitative features of $\mu$, in line with experimentally observed universality of mortality curves of different species. We average over $100$ networks with $1000$ simulations each; thus the lowest probability event we can resolve is of the order $\sim10^{-5}$, and fluctuations on that order is likely noise. (a) A higher damage probability $\gamma_0$ shifts the lifespan distribution and mortality curve to the right (b) Repair rate changes the plateau value $\mu_0$  (c) Increasing $N$ increases the slope of $\mu$ in the aging (Gompertz) regime. Only for large, complex networks do we find $\mu$ varying significantly with $t$; simple organisms do not age. 
(d) The initial damage causes a high infant mortality, but the damage is efficiently repaired soon after birth}
\end{figure*}

These two schemes are not altogether arbitrary. They can be justified using biological arguments and are bolstered by empirical evidence. SFN structures are known to accurately describe the organization of the metabolic networks of many different species \cite{jeong}. A simple evolutionary mechanism  \cite{stoltzfus} can give rise to a RN structure and account for the complexity of various biological structures via an entropic argument: Since the number of configurations in which components randomly depend on each other is overwhelmingly more numerous than the configurations in which all elements are fully independent, an initially-fully-independent collection of components will, statistically speaking, inevitably move towards random interdependence, provided that the fitness cost of introducing a dependency is negligible.

We assume that the aging of complex dependency networks are governed by the following three rules: (1) Every component in the organism must depend on at least one other node, and at least one other node must depend on it (i.e. all parts of the organism must be fully connected). (2) With certain fixed small probabilities the components can break (stop functioning) or be repaired (start functioning). (3) A node stops functioning if the majority of those on which it depends (providers) stop functioning, and cannot be repaired without a majority of its providers functioning.

Our simulations are based on the three rules listed above implemented as follows.  
\begin{enumerate}
\item Create a network model of an organism
\begin{enumerate}
\item  Begin with a single node, and $i=1$.
\item Introduce a new $(i+1)^{th}$ node and make it depend on any one of the pre-existing nodes $j\le i$ with probability $P(k_j)$, where $k_j$ is the degree of node $j$. For the neutral scheme $P(k_j)$ is taken to be uniform and independent of $k_j$, whereas for the non-neutral scheme $P(k_j)$ is taken proportional to $k_j$.
\item Make any existing node $j$ depend on the $(i+1)^{th}$ node with probability $P(k_j)$
\item Increment $i$ and repeat step b and c for $N-1$ steps.
\end{enumerate}
\item Age the resulting network model of an organism
\begin{enumerate}
\item Define the organism functional vector $\vec{\psi}(t)=\{x_1(t),x_2(t),\ldots,x_N(t)\}$ where every component $x_i$ can take either one of the values 1 (functional) or 0 (non functional). The vitality of the organism is defined as $\phi(t) = \Sigma_i x_i(t) / N$. Assign a value of 0 to a fraction $d$ of randomly selected nodes and 1 to the rest, corresponding to the initial damage of functionality in an organism. 
\item For all $i$, update $x_i=1$ to $x_i=0$ with probability $\gamma_0$, flip $x_i=0$ to $x_i=1$ with probability $\gamma_1$, and do nothing with probability $1-\gamma_0-\gamma_1$.
\item Break a node if the majority of nodes on which it depends are broken. Recursively repeat until no additional node breaks.
\item Set $\vec{\psi}(t+1)$ to the outcome of step (c).
\item Increment t and repeat (b-d) until all nodes are broken (i.e. $\sum_i x_i(t)=0)$.  
\end{enumerate}
\end{enumerate}
\begin{figure}
\centering
\includegraphics[width=3.0in]{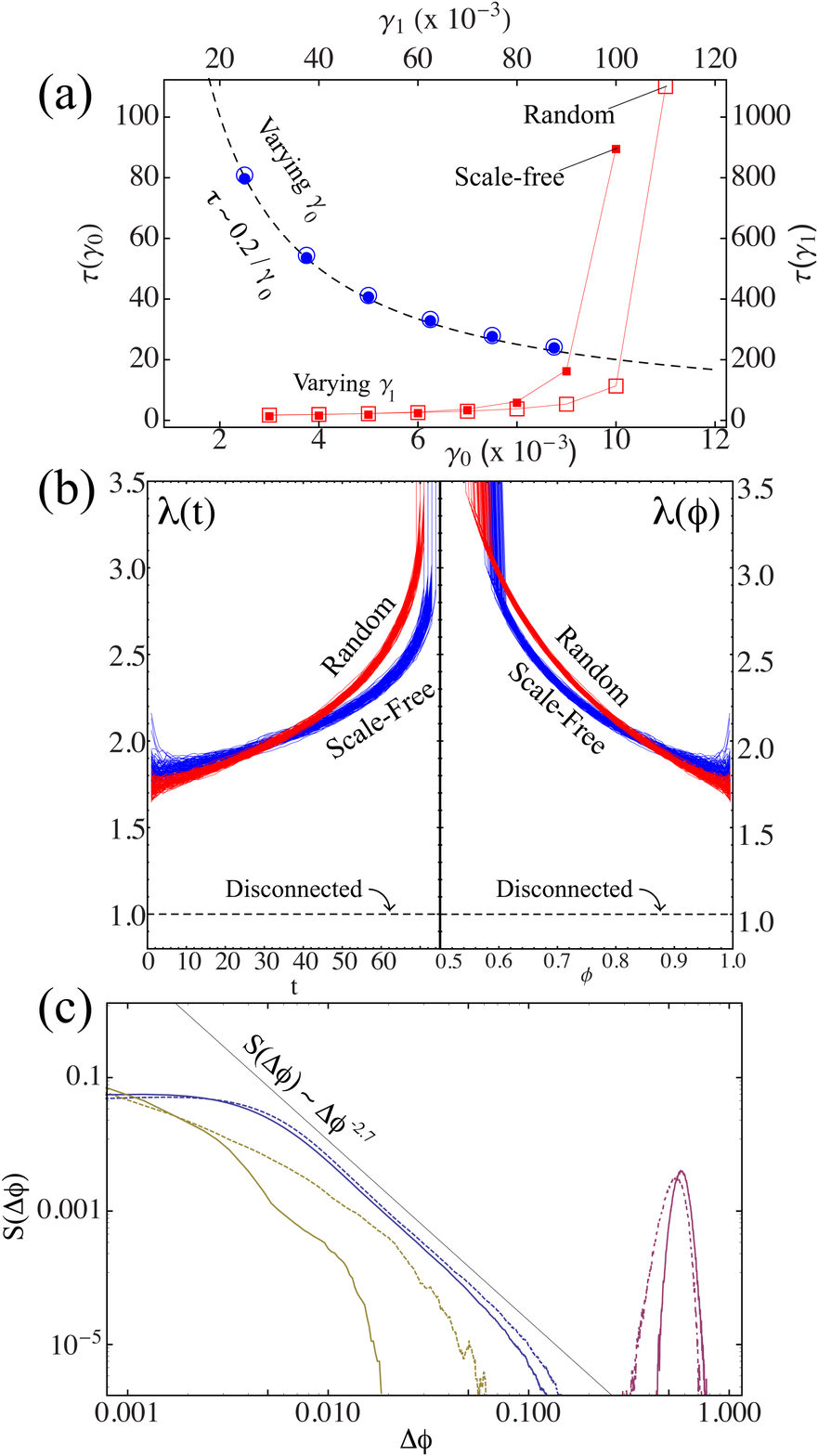}
\caption{{\bf Lifetime, interdependence and event distribution.} {\scriptsize (a) Average lifetime $\langle \tau\rangle$ versus damage rate $\gamma_0$ (blue circles) and repair rate $\gamma_1$ (red squares) for scale-free networks (filled markers) and random networks (open markers) with $N=2500$. Dashed lines mark $\tau=0.20/\gamma_0$ for both random and scale free networks. $\gamma_0=0.0065$ is kept constant for both red curves. There seems to be a critical repair rate $\gamma^*$ for which expected lifespan diverges. Note the remarkable independence of the curves with respect to network structure. (b) $\lambda(t)=\log[\phi(t)]/\gamma_0 t$ as a function of $t$ (left) and
$\phi$ (right) for 100 trajectories on the scale free (blue) and
random (red) networks with $N=10^6$ pictured in Fig[1] ($\gamma_1=0$).  The
interdependence parameter $\lambda$ varies strongly with both $t$ and
$\phi$, and roughly doubles as more damage is accumulated, until the sudden collapse depicted in Fig[1] leads to a diverging interdependence.  Scale free networks show
a relatively large variation in $\lambda$ for short times, but rapidly
converge on a monotonic increase as the network accumulates damage.
Interestingly, scale free networks begin with a larger value of
$\lambda$, but random networks become more interdependent rapidly.
The interdependence of a set of disconnected nodes (completely
independent) have $\lambda_{0}=1$ shown in the dashed line. (c) Probability $S$ that $\phi$ drops by $\Delta\phi$ before the largest drop happens (blue) of the largest drop itself (purple) and after the largest drop happens (yellow). Scale free (solid) and random (dashed) networks show a remarkable similarity. The largest drop distribution of both random and scale free networks obey a power law with exponent $-2.7$, (black line marks slope). Note that ``disease'' (blue) is qualitatively different from ``death'' (purple). Simulation parameters are $\{N,\gamma_0,\gamma_1,d\}=\{2500, 0.0025, 0,0\}$.}}
\end{figure}

In order to study the network mortalities, we must define a time of ``death'' $\tau$, which could be chosen in multiple ways. We define a threshold $\eta=\phi(\tau) = 1\%$ below which an organism is defined dead \cite{foot1}. The effect of the threshold value in the vitality $\phi(t)$ that we use to define death does not affect our simulations for large $N$.


To establish the statistical properties of mortality in our network, we generate an ensemble of networks (organisms) and age them according to the above rules. In addition to tracking the vitality of the network characterized by $\phi(t)$, we also determine the fraction of networks (not components) $s(t)$ that remain alive at time $t$ so that the time dependent mortality rate is
\[\mu(t)=-[s(t+1)-s(t)]/s(t)\]  
We also track the degree of interdependence within a given network (organism), characterized by the ratio 
\[\lambda[\phi(t)]= \log [\phi(t)]/\log[\phi_{0}(t)],\]
where $\phi_{0}=\mbox{exp}\{(-\gamma_0+\gamma_1)t\}$ is the expectation value of the vitality of an identical size network with all dependency edges removed. Thus $\lambda$ quantifies how much more often a network dies in comparison with one that has no interdependent components.

Finally, in order to analyze the magnitudes of functionality loss we consider the probability distribution $S[\Delta\phi]$ of event sizes $\Delta\phi$ (i.e. changes in $\phi$). Each event represents an individual disease or recovery, the final one of which is death.

\section{Results}
In our simulations a typical organism starts its life by a slow decay of $\phi$ at a rate of $\langle a \rangle \gamma_0$, where the dimensionless number $\langle a\rangle =1.80$ for scale free networks and $\langle a\rangle =1.75$  for random networks. As an increasing number of nodes die, the system approaches a critical vitality $\phi(\tau)=\phi_c$ when all live nodes suddenly collapse. Typical trajectories for both network topologies as well as the values of $\langle a\rangle$ and $\phi_c$ 
are shown in Fig[1]. 

We observe that even if $\gamma_1$ is set equal to $\gamma_0$ the system decays steadily despite the seeming reversibility in dynamics. This is because while any live node can break, not all the dead nodes will have the sufficient number of live providers to sustain a repair. 

A number of qualitative features of the mortality curves predicted by our model are independent of the range of parameters we explore and the network structures we chose. In particular, we see that the following features are generic (1) There is a slightly higher infant death rate proportional to initial damage $d$ followed by a sudden drop in death rate in early childhood 
(2) There is an exponential increase in the mortality following this initial blip, consistent with the Gompertz Law and (3) There is a final plateau in late-life mortality. In our simulations aging decelerates in late life and in any case does not increase beyond a critical $\mu_0$, the value of which is modulated by $\gamma_1$ (Fig[3b]). This is particularly clear in Fig[2], where we compare the mortality curves generated by our digital populations to that of a variety of organisms,  \emph{C. elegans, Drosophila}, Medflies, Beetles, Mice, Himalayan Goats (Tahr), using data compiled from \cite{horiuchi, caughley,vaupel}, and see reasonable agreement between simulation and data.

Since complex interdependence in networks is not exclusive to living organisms, we also fit the empirical aging curves to our model, of two kind of automobiles, the 1980 Toyota and 1980 Chevrolet (see Supplementary Information, FigS.1) as gathered from \cite{vaupel} .

An empirical analysis of historical human mortality data has shown that plotting the mortality rate $\mu(t_1)$ at some age $t_1$ against that at infancy $\mu(t_0)$ gives a ``universal'' curve that nearly overlaps for many different societies and historical periods \cite{azbel}. In other words, for an arbitrary collection of people (presumably with different damage /repair rates and initial conditions), the mortality rate at any two ages are correlated, with correlation coefficient $\rho\sim1$. In our simulations we vary $\gamma_0,\gamma_1,d$ while keeping $N$ constant and determine correlation coefficients between mortality rates from ages $t0$  to $t0+5$ and from $t_1$ to $t_1+5$. For $\{t_0,t_1\}=\{0,20\},\{10,20\}$ and $\{15,20\}$  is correlated by $\rho\sim0.8, 0.98$ and $0.99$. These correlation values are nearly identical for scale free and random networks. For details, see supplementary information and fig.S.6.


The effects of the system parameters on the mortality rate for both scale free and random networks can be summarized as follows: Increasing $\gamma_0$ shifts $\mu(t)$ left (Fig[3a]); increasing $\gamma_1$ decreases the value of the value of $\mu_0=\mu(t\to\infty)$ at the late life plateau (Fig[3b]); increasing $N$ increases the slope of $\mu$ in the aging (Gompertz) regime (Fig[3c]); in other words, larger systems age rapidly and suddenly, while small systems with few components are virtually non-aging. Increasing the initial damage $d$ simply elevates the initial (infant) mortality rate (Fig[3d]).  Our simulations yield a negative correlation between damage and \emph{increase in} mortality rate; the high initial damage populations age slower than the low initial damage populations to eventually converge to the same $\mu_0$ consistent with Strehler-Mildvan correlation law \cite{strehler}.

The qualitative dependence of average lifetime on damage and repair rate is as intuitively expected  (Fig[4a]). Quantitatively, when $\gamma_1=0$ the average lifespan perfectly fits the curve $\langle\tau\rangle=\beta/\gamma_0$, with the same value of $\beta=0.2$ fitting both scale free and random networks. Curiously, lifespan is very weakly dependent on the repair rate $\gamma_1$ for smaller values until it rapidly diverges at a critical value of $\gamma_1^*$. This makes one wonder why most species are not immortal, but Fig[4a] suggests why:  For small $\gamma_1$, $\tau(\gamma_1)$ is nearly constant, $(1/\tau)\partial\tau/\partial\gamma_1\ll1$. However, if the cost of repairing nodes increases with the probability of repair $\gamma_1$, increasing $\gamma_1$ becomes evolutionarily nonviable due to the weak dependence of $\tau$ over a wide range of $\gamma_1$. That being said, apart from this high evolutionary cost (and the slim chance of a massive statistical fluctuation of the order $\gamma_0^N$ which obviously vanishes for $N\to\infty$) there is no mechanism in our theory that prevents a high-repair species from being immortal in the thermodynamic limit.


The dependency coefficient $\lambda[\phi(t)]$ gradually increases as our simulated organisms grow older, and diverges just before death (Fig[4b]). The functional form of $\lambda[\phi(t)]$ is qualitatively similar for SFN and RN.

We intuitively expect the onset of ``death'' to differ drastically from the early aging process (referred to as ``disease'').  This difference will be reflected in the distribution of the number of living nodes that die in a particular time step ($\Delta \phi=\phi(t)-\phi(t-1))$.  
In order to tell whether death is just ``the last disease'' or a qualitatively different phenomenon, we analyze the distribution $S(\Delta\phi)$ of event sizes before, after and \emph{of} the largest drop and notice that the latter is qualitatively as well as quantitatively very different from the former two. Death and disease occupy an entirely different region of the event spectrum (Fig[4c]). What is even more remarkable is that the disease distribution for both evolutionary schemes RN and SFN are quantitatively similar over a wide range of $\Delta \phi$, and obey a power law $S(\Delta\phi)\sim1/\Delta\phi^{2.7}$. We do not have an explanation for this striking similarity, nor the value of the critical exponent.

We cautiously note that while a set of  four parameters uniquely determines $\mu(t)$, the converse is not true. In particular the characteristic features of $\mu$ (as defined by the initial slope, the plateau value $\mu_0$, average lifespan $\tau$ and crossover time from an aging to non-aging regime) can be kept ``similar'' by holding $\gamma_0$ constant while increasing $N$, $\gamma_{1}$ and $d$ simultaneously (see Supplementary Information and Fig.S.4). Of course, there is no a priori reason why two species with different attributes must \emph{necessarily} have different aging curves.

\section{Theory}

We now aim to obtain the values of initial decay rates $\langle a\rangle\gamma_0$, the critical vitality $\phi_c$ and understand why dependency networks collapse suddenly. On the way, we also hope to understand why these quantities are so similar for both scale free and random networks, and determine the origin of the Gompertz-like  law.

When the system is far from collapse, the probability that two providers of a single node dying at once $\mathcal{O}[\gamma_0^2]$ is negligible compared to that of a single provider dying $\mathcal{O}[\gamma_0]$. Then the total probability $p_0$ that a node dies is $\gamma_0$ plus the probability that the last vital provider of a node dies. If $m(\phi)$ is the probability that a node is left with one last vital provider, we can self-consistently evaluate $p_0$
\begin{align}\label{0}
p_0=\gamma_0+m(\phi)p_0(1-\gamma_0).
\end{align}
In a single step associated with the aging of the network,  the probability that a node is repaired is $p_1=h(\phi)\gamma_1$, where $h(\phi)$ is the probability that a node has at least the minimum number of providers required to function. Then, the change in the fraction of nodes that are alive, is given by  
\begin{align}\label{1}
 \Delta \phi&=p_0\phi-p_1(1-\phi)\nonumber\\
&=-\frac{\gamma_0\phi}{1-m(\phi)(1-\gamma_0)}+\gamma_1h(\phi)(1-\phi)
\end{align}
where we have used the expression for $p_0$ as obtained from (\ref{0}). From (\ref{1}), we see the origin of  the catastrophic (and universal) nature of death. For any arbitrary ``fully connected'' network and monotonically decreasing $\phi(t)$, the vital fraction $m(\phi)$ must always start from a finite value in the domain $[0,1]$  and increases towards unity as $\phi$ decreases, inevitably to cause the first term to dominate the second ($\gamma_0\ll1$), and thus leading to a sudden drop in the expected vitality. This is true in general, although the detailed form of the evolution of $\phi$ depends on the fraction $m(\phi)$ of vital providers and repairable fraction $h(\phi)$  that will vary for different network structures.

Equation (\ref{1}) also indicates an asymptote in longevity for large repair rates, as seen in Fig[4a]. If we set $\Delta\phi=0$, we find that the system lives indefinitely when the repair rate is set to $\gamma_1=\gamma_1^*$ where
\begin{align}\label{1.5}
\gamma_1^*=\frac{\gamma_0\phi^*}{h(\phi^*)(1-\phi^*)[1-m(\phi^*)(1-\gamma_0)]}
\end{align}
for any given $\phi^* \in [\phi_c,1]$, i.e. for a vitality larger than the critical vitality $\phi_c$. In this case the system damage increases while the vitality decreases until it reaches $\phi=\phi^*$, but maintains that damage forever. Of course, (\ref{1}) governs the \emph{expectation value} of $\phi(t)$ which is  the actual value only in the thermodynamic limit $N\to\infty$. Thus, only a system of infinite size that satisfies (\ref{1.5}) can live indefinitely, since a finite system will die  (at least) with probability $\gamma_0^{N(\phi(t)-\phi_c)}$ due to statistical fluctuations.

In general, it is a non-trivial task to obtain the exact forms of $m(\phi)$ and $h(\phi)$ and thence the average lifetime, the critical damage fraction etc.  However we can obtain  the initial slope $\langle a\rangle\gamma_0=p_0|_{t=0}$ with which the vitality decreases, and the critical vitality $\phi_c$ at which the whole system collapses (see dashed lines in Fig[1]) for the case $\gamma_1=0$: Let the probability that a node with $k$ providers die be $\sigma^{(k)}$. Then we can recursively obtain $\sigma^{(1)}$ in terms of the others \cite{foot2},
\begin{align}\label{2}
\sigma^{(1)}=\gamma_0+P(1,1)\sigma^{(1)}+P(1,2)\sigma^{(2)}+P(1,3)\sigma^{(3)}+\ldots
\end{align}
where $P(1,i)$ is the probability that the provider of a degree-1 node has $i$ providers. The first term corresponds to the probability that a node dies independent of its connectivity.  

Since we neglect probabilities of order $\mathcal{O}[\gamma_0^2]$, initially only degree-1 nodes can be killed by the death of their providers. Thus substituting $\sigma^{(k)}=\gamma_0$ for all k apart from $k=1$, and using $\sum_i P(1,i)=P(1)$ we can obtain  from (\ref{2}) the initial probability that a degree-1 node dies,
\begin{align}
\sigma^{(1)}=\frac{(2-P(1,1))\gamma_0}{1-P(1,1)}
\end{align}
to find the expectation value of the initial slope we must average over the damage rate of all degrees, including $\sigma^{(k)}=\gamma_0$ for $k>1$
\begin{align}\label{slope}
\langle a\rangle\gamma_0 =\left.\sum_k P(k)\sigma^{(k)}\right|_{t=0}= \gamma_0\left(1+\frac{P(1)}{1-P(1,1)}\right)
\end{align}
Upon substituting the numerical values of  $P(1)$ and $P(1,1)$ for the networks we evolved, we obtain  $\langle a\rangle=1.75$ for the neutral scheme (RN) and $\langle a\rangle=1.80$ for non-neutral (SFN) scheme, i.e.  only a $\sim2.8\%$ difference between two topologies. These initial slopes are consistent with our aging simulations (Fig[1]).

To estimate the critical vitality $\phi_c$, we start by asking what the smallest value of $\gamma_0$ must be in order to kill any network in just one step. If there exist a critical vitality $\phi_c$, then a death rate of $\gamma_0=1-\phi_c$ would kill the network in one step. Making this substitution and letting $\sigma^{(i)}\to1$ for all $i$ in (\ref{2}) yields a simple but interesting result,
\begin{align}\label{crit}
\phi_c=P(1).
\end{align} 
For the networks we evolved, P(1) is $0.5$ for the neutral scheme and $0.6$ for the non-neutral scheme. These values are consistent with the critical vitalities we observed in our aging simulations (Fig[1]).

Having estimated the average damage rate and the critical vitality of the network, we now consider the nature of lifespan distributions. The probability of network survival $s(t)$ is equal to the probability of $\delta\phi$ being not greater than $\phi-\phi_c$. That is, $s(t)=1-\mbox{Prob}(\Delta\phi>\phi-\phi_c)\nonumber$. Since each node dies with probability $p_0$,
\begin{align}\label{3}
s(t)=1-\sum_{k=0}^{N\phi_c}\binom{N\phi}{N(\phi-\phi_c)+k}p_0^{N(\phi-\phi_c)+k}(1-p_0)^{N\phi_c-k} 
\end{align}
Using the aging rate of the network established earlier, i.e. $1.75\gamma_0<p_0<1$ regardless of the network topology or size, i.e. $p_0$ is independent of both these parameters. Thus, in the limit $N\to\infty$ the probability of death $1-s(t)$ simply becomes a unit step function. On the other hand for finite $N$, the step function softens, and we empirically interpret the rapid transition from $s=1$ to $s=0$ as aging. We see that by passing from finite size to infinite size, we also pass from the stochastic to deterministic case, and from gradual aging to ``instant aging''.

Finally, we consider the sharpness of the transition from $s=1$ to $s=0$ to see if there is any relation between our results and the classical Gompertz law for mortality. Evaluating (\ref{3}) exactly is nontrivial, since, strictly speaking, all quantities in (\ref{3}) except $N$ are random variables. However we can substitute the average (early-time) value of $\phi(t)=e^{-p_0 t}$ in (\ref{3}); this is done at the cost of narrowing the longevity distributions but hopefully not their functional forms. The mortality rate $\mu(t)=-\partial_t s/s$ obtained this way is plotted in supplemental Fig.S.2 and agrees well with the empirical Gompertz law, which states that $\log\mu(t)$ increases linearly in time.

\section{Discussion}

We have built on a rough similarity between large networks and complex organisms (and machines) to create a minimal model of aging. It is thus important to be self-critical here by comparing our study both with reality and with previous attempts. Any theory that aims to account for a phenomenon as universal as aging spanning both animate and inanimate objects needs to be robust, i.e. it should not have such strong assumptions and results that strongly depend on system details. The weak sensitivity of our outcomes to the details of biologically justifiable dependency network structures (Fig[1,2,3,4]) seems to satisfy this requirement. However, as already pointed out, we have four parameters in our model - network size, rate of damage and repair, and the initial damage, in addition to the fraction of dependency that determines if a node dies or not (1/2 in our case). It seems unlikely that any model could be simpler, and while this is sufficient to fit a variety of mortality curves in organisms and machines, the set of these parameters is not unique for an organism. Clearly, therefore, we need further constraints to make our model even more robust, and some of these may come from mechanistic limits on the parameter values. 

Comparing our analysis with earlier models, we note that a classical benchmark is provided by the application of  reliability analysis to biological aging \cite{gavrilov, laird}, which investigates the robustness of systems in which cells are connected in a parallel-series circuit (i.e. an organ lives as long as one of its cells remains, and the organism dies as soon as one of its organs fails). While this picture can successfully explain some qualitative features of mortality curves, a few puzzles remain. First, it cannot account for the higher infant mortality followed by a drop (in contrast see Fig[3d] and second row of Fig.S.6) despite introducing an initial damage. Secondly, the plateau value of $\mu_0=\lim_{t\to\infty}\mu(t)$ implied by the model of \cite{gavrilov}  regardless of the number of cells per organ, is equal to $1-(1-\gamma_0)^m$, where $m$ is the total number of organs. This implies that any link to actual experimental values of $\mu_0$ for an organism such as \emph{C. elegans}, with say, $m\sim10$ organs, requires about ten percent of all cells die per day, an unreasonably large fraction that contradicts experiment. The final puzzle in \cite{gavrilov, laird} is, the predicted functional form of $\mu(t)$ depends very sensitively on a specific assumption of how the reliability circuit is built. Specifically, the functional form of the probability distribution describing the number of ``redundant cells per organ'' which by itself seems to have no a priori or experimental justification.  In contrast, our analysis of random and scale free networks
is able to explain the initial increase in mortality, is consistent with ``low'' rates of cell deaths seen in experiments, and is surprisingly independent of connectivity distributions or individual realizations of networks.

Our analysis also differs significantly from earlier theoretical investigations of network  failure \cite{barabasi, stanley1, cohen, stanley2} in which nodes are simply removed one by one (systematically or randomly) until networks get fragmented. In these approaches the nodes do not influence the performance of one other, and are not allowed to be repaired. Curiously, the lack of interactions in these models lead to fundamentally different fragmentation dynamics in scale free and random networks. In contrast the survival curves we observe are remarkably independent of the network topology (Fig[3,4]). The strong interactions between components may be the reason behind the strong similarity of mortality curves among so many organisms; a conclusion that bears similarity to that of another strong-interaction model that has been proposed to explain electrical gird failures \cite{stanley2}. Much still remains to be done in understanding how the form of these interactions leads to differences or similarities in the dynamics. 

Our study has focused on the dynamics of networks that age as a consequence of interdependency, and thus leads naturally to the question of how this might be controlled. Since the repair rate, and perhaps the damage rate (to a lesser extent) are experimentally controllable, one might ask if it is possible to vary their temporal character while keeping their average constant. Are there optimal strategies for repair - either in the time domain or in space (i.e. looking at nodes with varying connectivity)? For example, Fig[4a] shows that if the system is repaired uniformly, the degree of repair does not make a significant difference for $\gamma_1<\gamma_1^*$. It would be very useful to know if and how a (temporal or spatial) non-uniform repair strategy improves lifespan. In networks that are dynamically heterogeneous, we may ask what would be the consequences of differential damage and repair in a network with highly variable turnover - e.g. a network with  tissues like the skin or gut that have high damage and repair rates, and the brain which has a low damage and repair rate? At the level of ecology and colonies, we might ask how does the aging dynamics of a dependency network change when a system consists of parts with aging rates comparable to that of the whole system?  We hope that our minimal model may be used to study some of these questions.


\begin{acknowledgments} DCV thanks Anthony J. Leggett for his support, and to Pinar Zorlutuna for many stimulating discussions. This work was partly supported by grants NSF-DMR-03-50842, NSF-DMR09-06921, the Wyss Institute for Biologically Inspired Engineering, the Harvard Kavli Institute for Bio-nano Science and Technology, and the MacArthur Foundation. 
\end{acknowledgments}

\appendix*
\section{Supplemental Information}
\setcounter{figure}{0} \renewcommand{\thefigure}{S.\arabic{figure}}
Here we discuss in more detail various technical points surrounding theory, simulations, and fits.\\

1. Although we have discussed aging in an evolutionary setting, specialized interdependence of components is not exclusive to living organisms. To demonstrate this point we fit our model to the empiric mortality curves of 1980 Toyota and 1980 Chevrolet obtained from \cite{vaupel}. We were able to obtain a reasonable agreement using identical values of $N$ and $d$ but slightly varying damage and repair rates for the two cars (Fig S.1).\\

\begin{figure}[h!]
\includegraphics[width=1.66in]{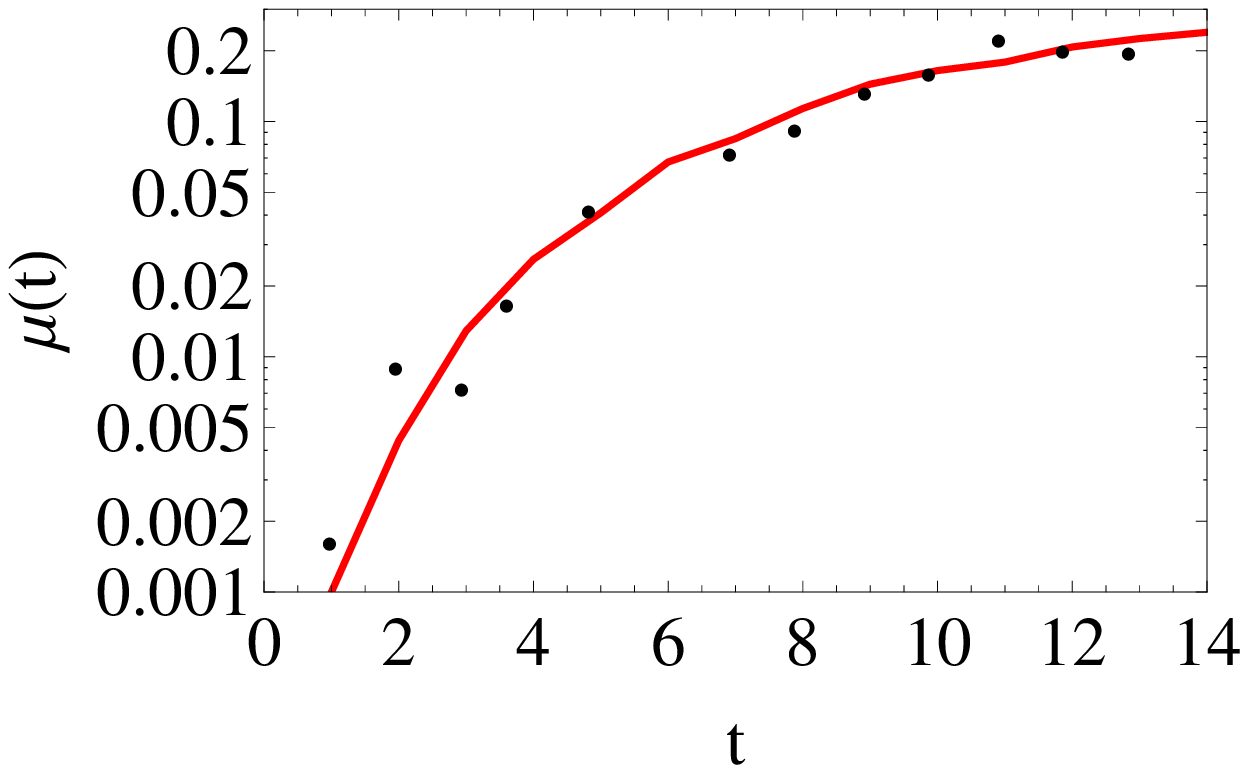}
\includegraphics[width=1.66in]{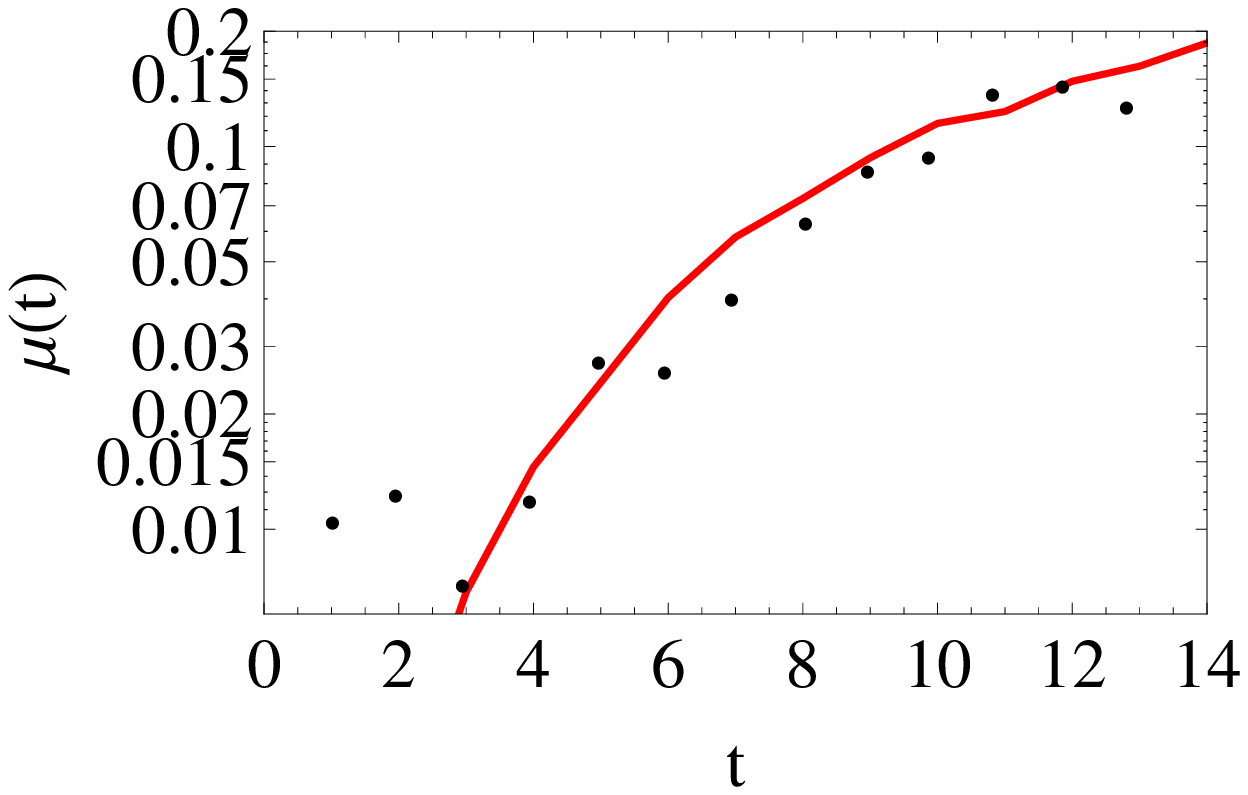}
\caption{{\bf Empiric mortality rates of cars (black) fit to present theory (red).} The data for 1980 Toyota (left) and 1980 Chevrolet (right) from \cite{vaupel} is fitted with $\{N,\gamma_0,\gamma_1,d\}$= $\{200,0.023,0.023,0\}$ and $\{200,0.02,0.02,0\}$ respectively. The horizontal axis denotes years.}
\end{figure}

2. In order to recover the Gompertz Law analytically, we substitute $\phi(t)\approx e^{-a\gamma_0t}$ for $t\ll\tau$ in (8) and plot $\mu(t)=-(ds(t)/dt)/s(t)$ in Fig.S.2. By approximating the random variables $\phi(t)$ and $p_0$ by their average value we sharpen the lifetime distributions and hence steepen the mortality curve; however Gompertz's Law is still recovered for short times (Fig.S.2).\\

3. We have defined death as the time $\tau$ at which $\phi$ reaches a threshold value $1\%$, which may seem arbitrary. In order to determine how sensitive our results are to the choice of threshold $\eta$, we have analyzed the lifetimes of both network topologies as a function $\eta$. For a (small) network of $N=2500$ Fig.S.3 shows that the value of $\eta$ changes the lifetime less than 1\% for a SFN and less than $12$ for a RN. We find that $\eta$ dependence rapidly vanishes for $N>\mathcal{O}[10^3]$ for both network types and conclude that the precise value of $\eta$ is not important.\\

4. To demonstrate that our model parameters have individual predictive power we fit the empirical data of a mutant and wildtype of a fixed organism C.Elegans by fixing $N$,$\gamma_0$ and $d$ constant, but only varying the repair rate $\gamma_1$ (Fig S.4). The difference in fit parameters as noted in the legend of Fig[2] and Fig.S.4 stem from a discrepancy in the data presented in \cite{vaupel} and \cite{horiuchi} for the same organism.\\
\begin{figure}[h!]
\begin{center}
\includegraphics[width=3.0in]{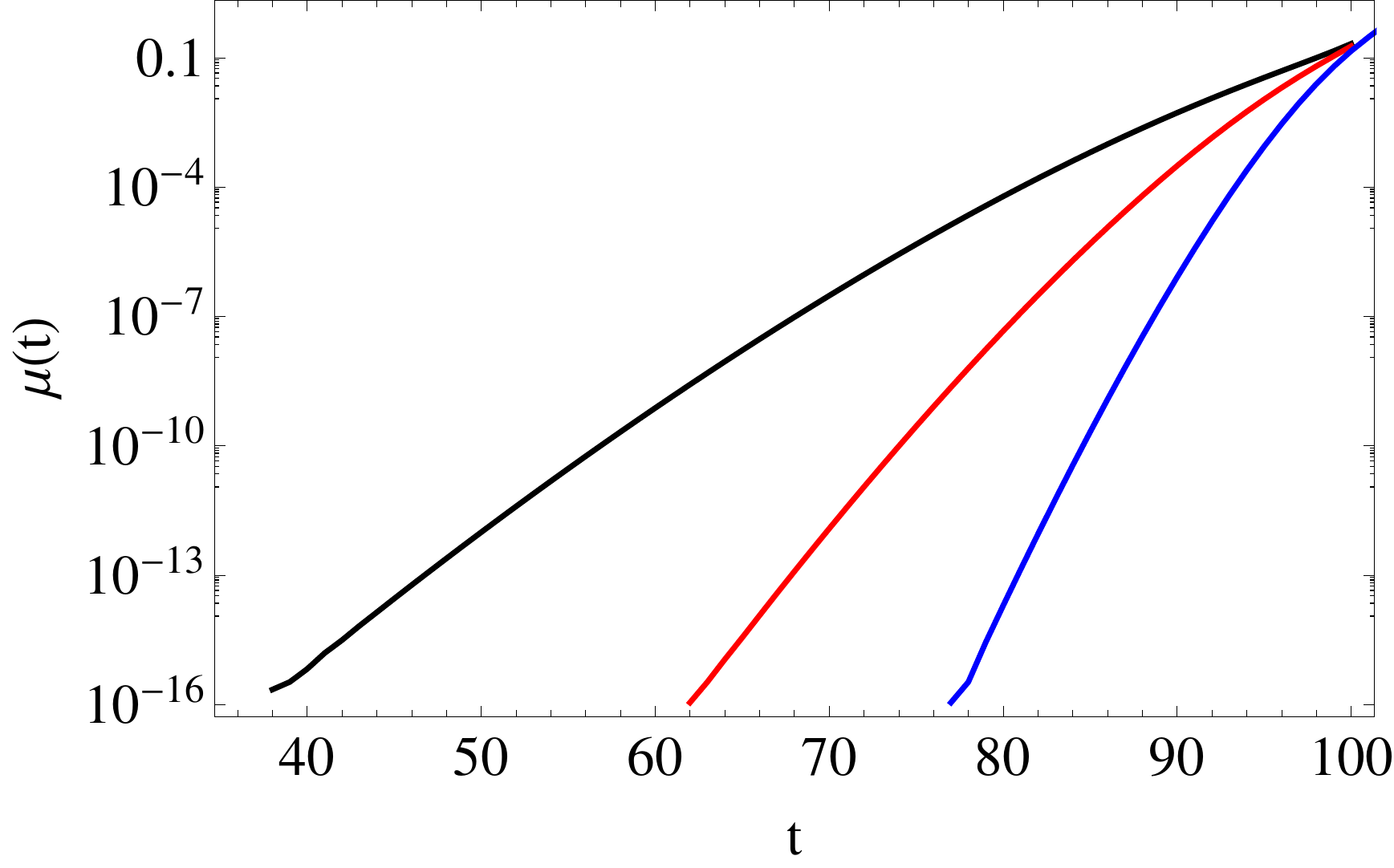}
\caption{{\bf Mortality vs time} obtained from (8) for $\phi\ll\phi_c$, $\gamma_0=0.0025$ and $N= 50$ (black) 100 (red) and 200 (blue) is in good qualitative agreement with our simulations and the empiric ``Gompertz law'' which states that $\log\mu(t)$ increases linearly with $t$ early in life.}  \end{center}
\end{figure}

\begin{figure}\begin{center}
\includegraphics[width=3.6in]{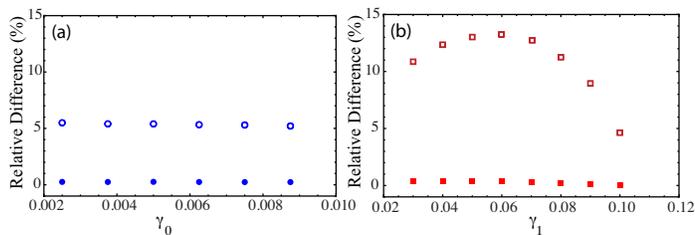}
\caption{{\bf Sensitivity of outcomes to the definition of death.} (a) The percent lifetime difference between choices $\eta=1\%$ and $50\%$ for scale free (filled circles) and random (empty circles) dependence networks, with varying $\gamma_0$ and $\gamma_1=0$.  There is a fairly constant $5\%$ difference independent of $\gamma_0$ for random networks, and below $1\%$ for scale free networks. (b) The percent lifetime difference between choices $\eta=1\%$ and $50\%$ for varying $\gamma_1$, with $\gamma_0=0.00625$.  Scale free networks continue to have below $1\%$ difference between the two threshold choices, while the variation is more significant for random networks with large repair rate $\gamma_1$. The relative difference decreases if the two thresholds $\eta$ are closer to one another. For both graphs $N=2500$, $d=0$.}\end{center}
\end{figure}

\begin{figure}
\begin{center}
\includegraphics[width=2.5in]{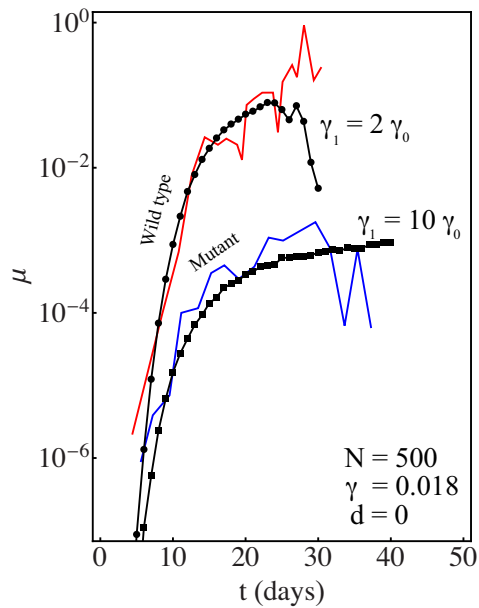}
\caption{{\bf Mortality Rate of Mutant and Wild-Type Nematodes.} We fit the data from Fig. 3E of \cite{vaupel} using our model.  The data includes a wild type {\em{C. Elegans}}, which is well fit by the parameters $N=500$, $\gamma_0=0.018$, $\gamma_1=2\gamma_0$, and $d=0$.
 The mortality curve for the {\em{Age-1}} mutant of the worm is well fit using the same values of $N$, $\gamma_0$, and $d$, but with $\gamma_1=10\gamma_0$ significantly increased.}
\end{center}
\end{figure}

5. To determine whether model parameters $\{N, \gamma_0,\gamma_1,d\}$ can be determined uniquely given experimental $\mu(t)$ data we performed simulations sweeping the parameter space $N\in \{50, 100, 250, 500, 700, 800, 900, 1000, 1500, 2000\}$,\linebreak  $\gamma_0\in \{2, 4, 6, 8, 10, 15, 20, 25, 30, 35\} \times 10^{-3}, \gamma_1\in \{1,2,\ldots,10\}\times \gamma_0, d\in\{0,0.5,.\ldots,10\}$ (percent) and checked if non-neighboring parameters give more similar $\mu(t)$ curves than neighboring ones (cf. below for details). For each set of parameters we generated 12 networks upon which 3000 simulations were performed, providing a reasonable level of confidence in the statistical accuracy of the simulations.  To quantitatively compare the simulation results, $\mu(t)$ is broken into four averaged characteristics:  The initial slope, the saturation point, the crossover time between the initial growth and saturation, and the observed lifetime (see Fig.S.5a). The threshold for similarity of the curve characteristics is determined by averaging over the differences in nearest neighbors in the $(N,\gamma_0,\gamma_1,d)$ parameter space (i.e. the $\sum_{k=1}^8 |\tau_{ref}-\tau_{S_k}|$, with the parameters in the simulation $S_k$ being different from the reference simulation in only one position, and a nearest neighbor). Thus, simulation outcomes are considered ``similar'' to a reference if they are not nearest neighbors (in parameter space) with the reference, and if the differences in all four characteristics simultaneously fall within the threshold variation. Fig.S.5b shows one such overlapping curve with $d=4.5$, $N=700$ and $\gamma_1=5\gamma_0$ (compared to $d=0\%$, $N=250$ and $\gamma_1=0$ for the reference simulation in black).  The inset of Fig.S.5b shows all 214 simulation parameters that yields mortality curves considered "similar" to the reference curve (about 0.9\% of all of the simulated parameters), and shows that a rather wide range of parameters may give qualitatively similar behavior in $\mu$ and $\tau$ (with the latter not shown).  It is interesting to note that $\gamma_0$ is the same for the reference curve and all overlapping curves, indicating that an empirically observed death rate $\gamma_0$ may be uniquely determined.\\

\begin{figure}
\includegraphics[width=3.6in]{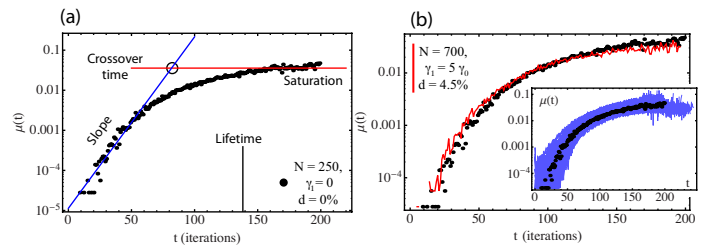}\begin{center}
\caption{{\bf Uniqueness of $\mu(t)$ as determined by initial slope, saturation value crossover time and average lifetime} (a)  A simulation is quantified in terms of four parameters:  The initial slope, the final saturation value, the crossover time, and the lifetime.  Shown is a reference simulation with $N=250$, $\gamma_0=0.002$, $\gamma_1=0$, and $d=0\%$  (b) The non-uniqueness of $\mu(t)$ as the parameters are varied.  In the main panel and the inset, the black points correspond to the reference simulation in (a).  The three red line in the main panel has $N=700$,
$\gamma_0=0.002$, $\gamma_1=5\gamma_0$, and $d=4.5\%$.  The average lifetime for the red curve is $\tau=133$, within 4.3\% of the lifetime of the reference curve.  There is moderate variation between the curves for small and large $t$, but it would be difficult to unambiguously differentiate between the two sets of parameters when fitting experimental data. The inset shows the same reference simulation (black points), along with all 214 sets of simulated parameters that satisfy the threshold criterion.  Each blue line has  $250 \le N\le 2000, 0\le\gamma_1/\gamma_0\le 9$, and $0\le d\le 8.5$, all with  $\gamma_0=0.002$.}\end{center}
\end{figure}

\begin{figure*}[ht!]
\includegraphics[width=1.9in]{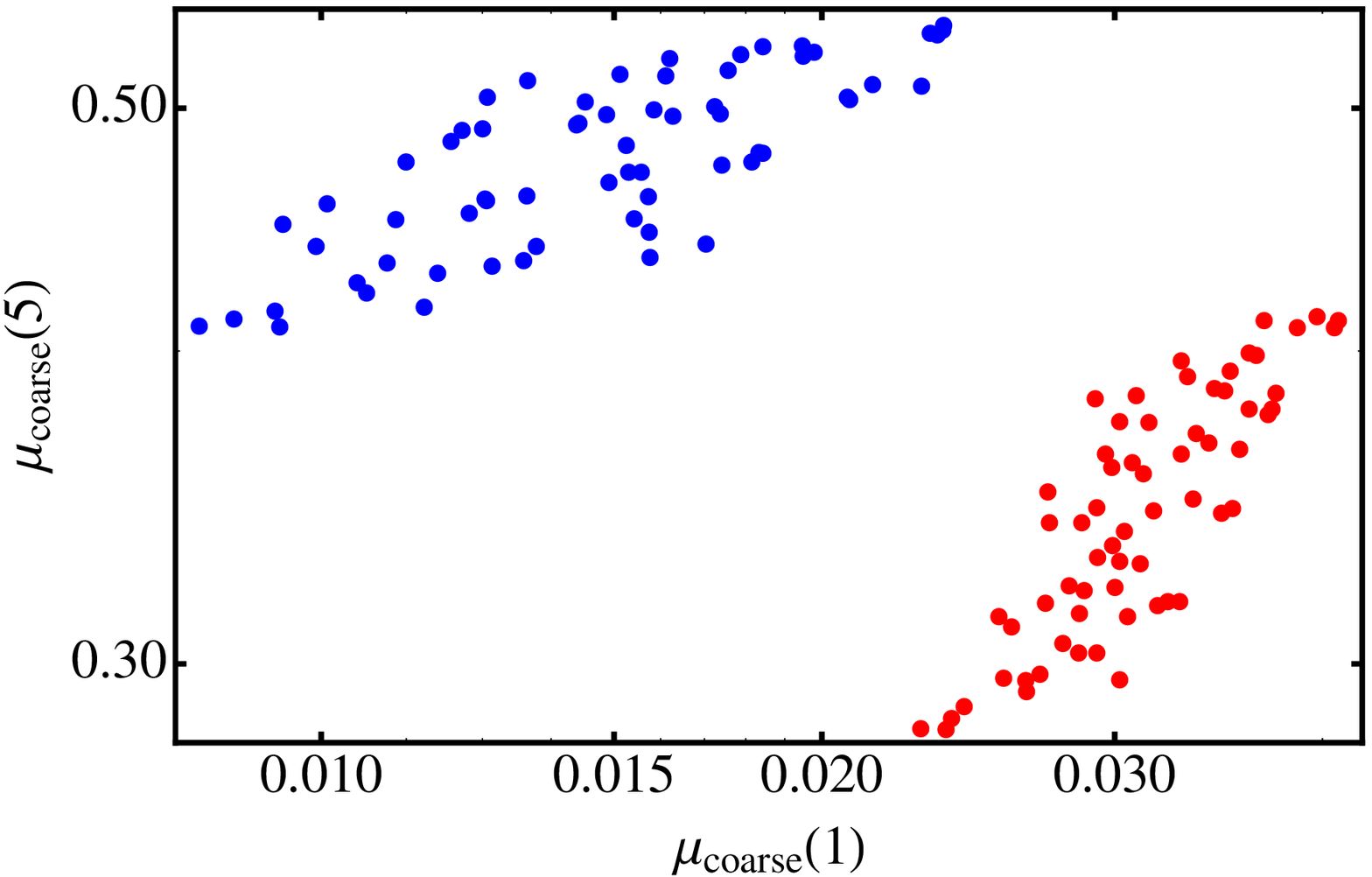}
\includegraphics[width=1.9in]{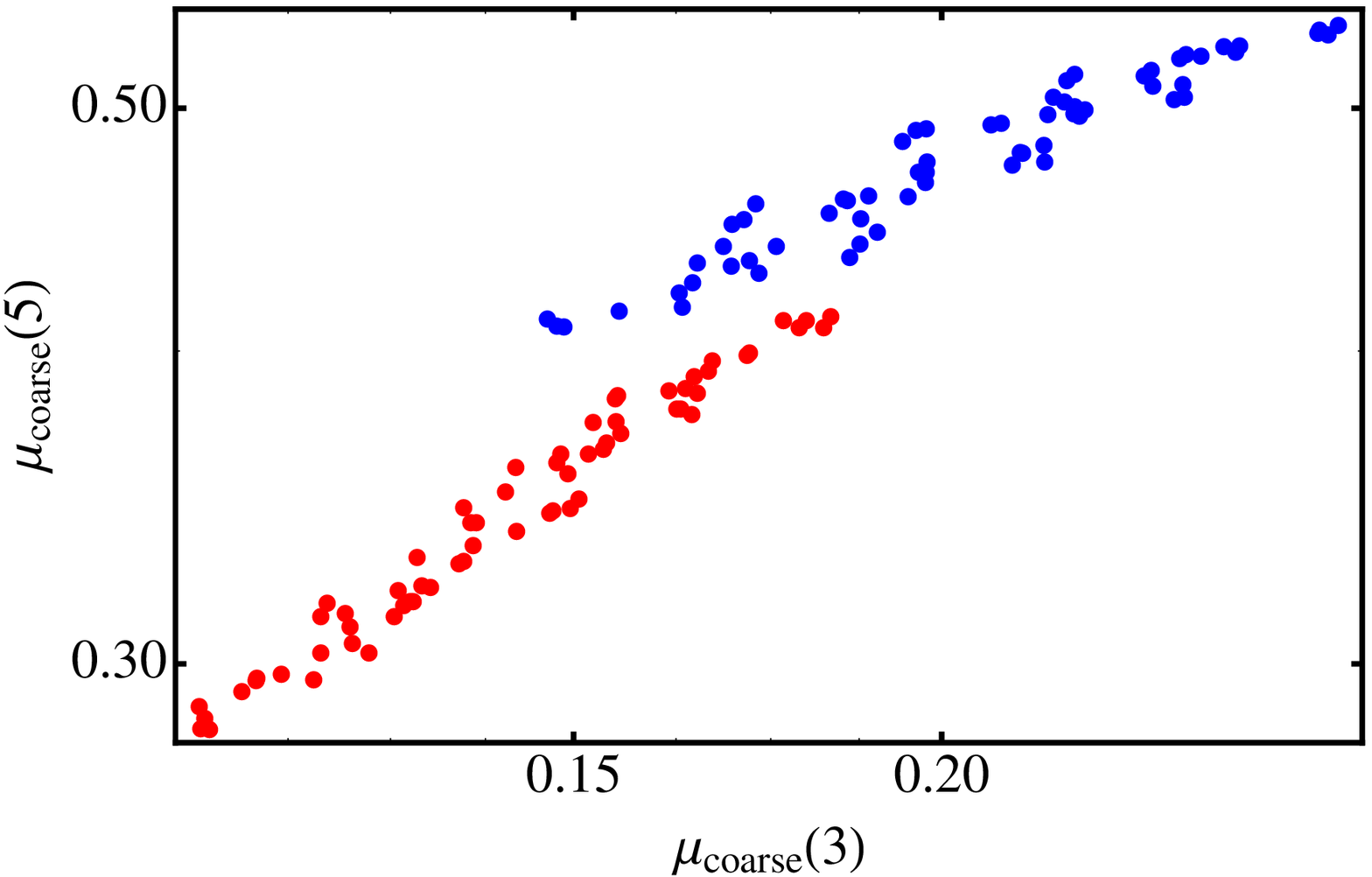}
\includegraphics[width=1.9in]{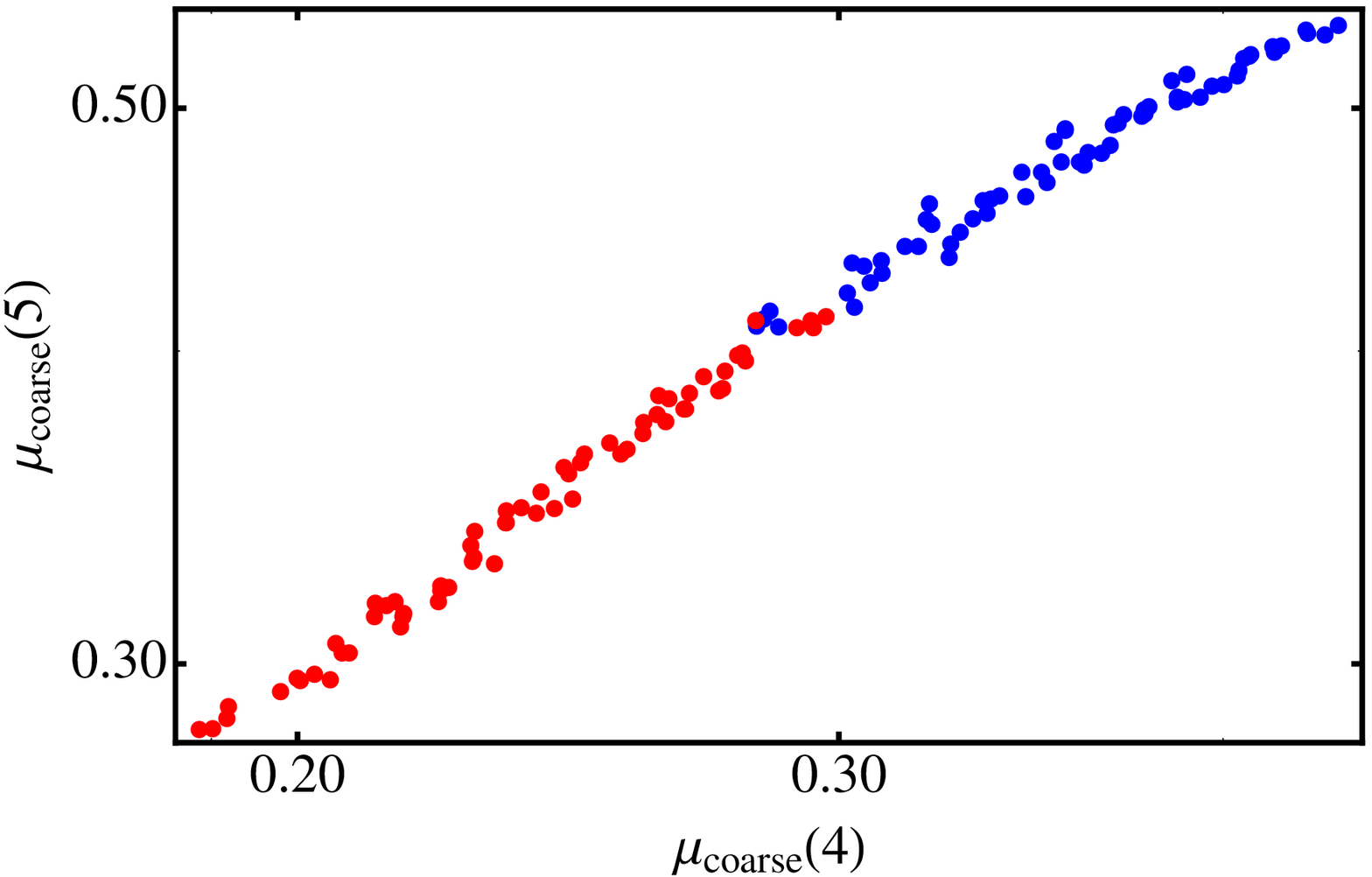}\\
\includegraphics[width=2.1in]{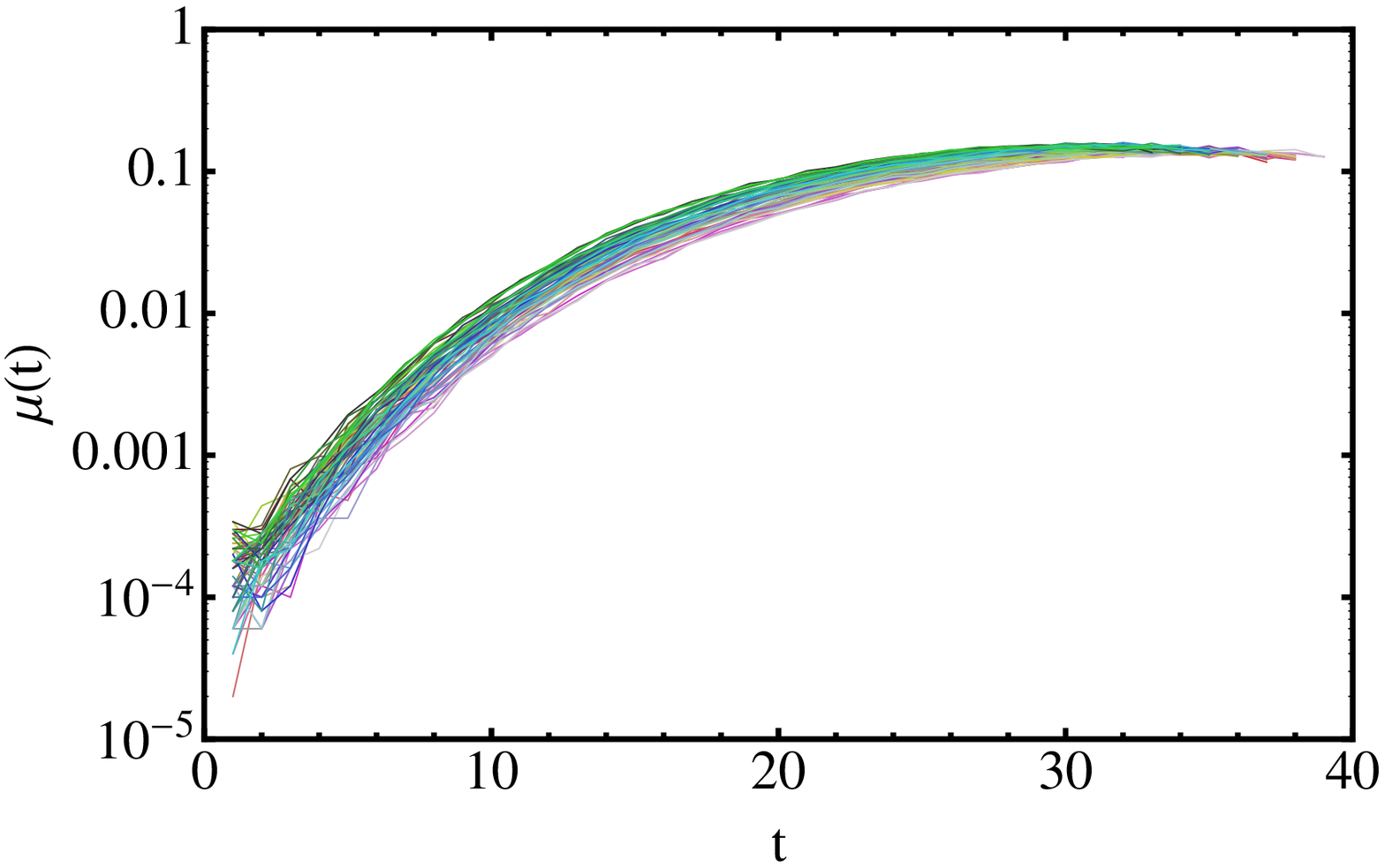}
\includegraphics[width=2.1in]{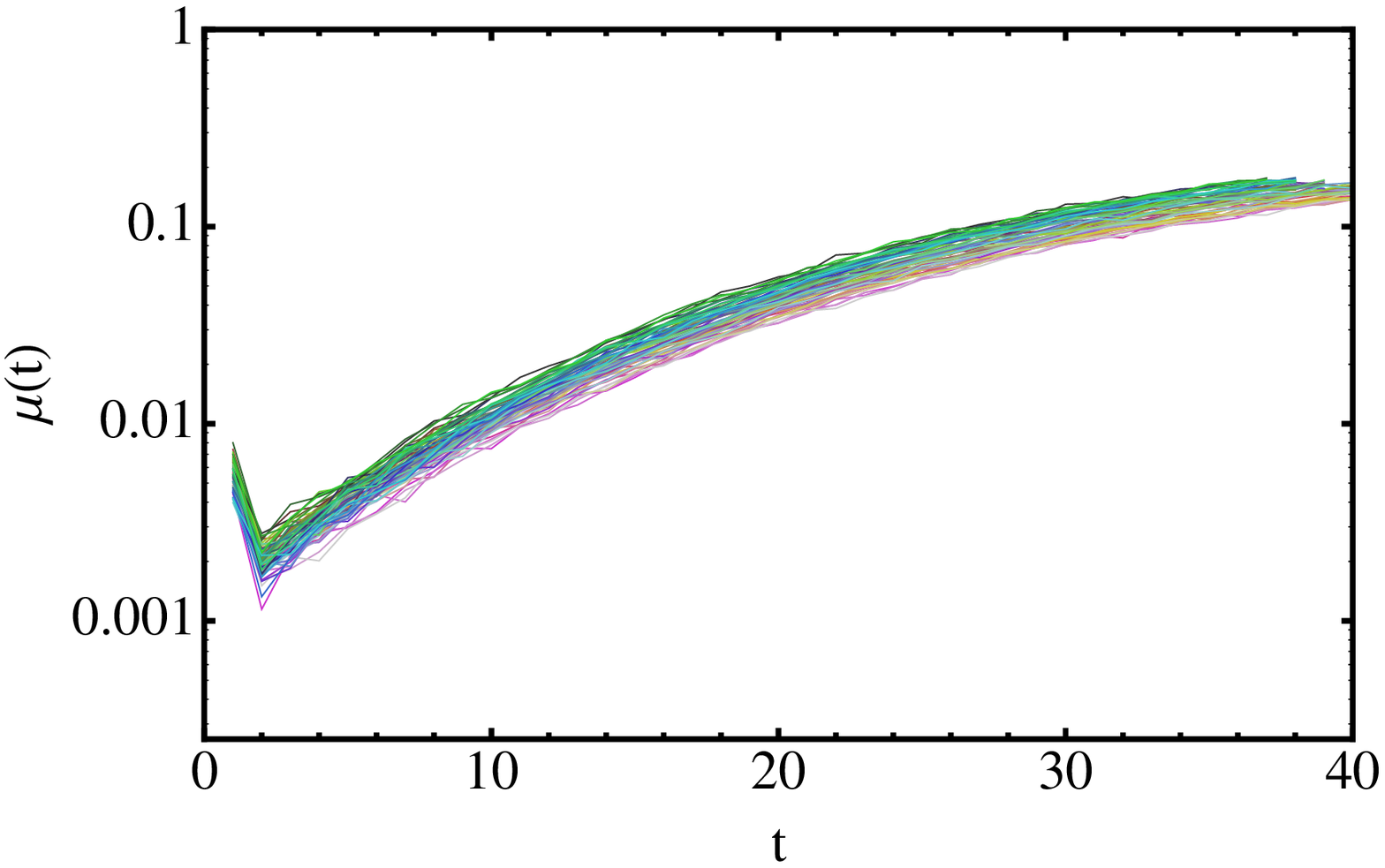}
\caption{{\bf Testing for universality in mortality patterns.} Dependence of old age mortality $\mu(5)=\mbox{Prob}(20<\tau<25)$ on younger age mortality $\mu(i)=\mbox{Prob}(5i-5<\tau<5i)$ for $i=1$ (top left), $3$ (top center), $4$ (top right) for a range of network types (RN blue, SFN red) and system parameters, $\gamma_0=\{0.0026,0.0027,0.0028,0.0029\}$, $\gamma_1=\{0.0026,0.0027,0.0028,0.0029\}$, $d=\{12.1,12.2,12.3,12.4\}$ ($\tau$ is the time of death). The mortality curves generated by this range of parameters is displayed in the bottom row, for RN (left) and SFN (right). While the universal (yet species-specific) trend observed in \cite{azbel} between $\mu(i)$ and $\mu(i+j)$ (for fixed j) is qualitatively present in our model, the trend vanishes for large enough $j$.
}
\end{figure*}
6. It has been empirically observed that the value of $\mu(t_0)$ at age $t_0$ correlates very strongly with $\mu(t_1)$ at age $t_1$ for a wide variety of societies and historical periods \cite{azbel}. To test whether our model yields this empiric ``universality'', we plot $\mu(t_0)$ versus $\mu(t_1)$ for a varying range of $\gamma_0,\gamma_1$, $d$ and network types. While the trend is less impressive for very widely separated $t_0, t1$, our model does yield a correlation between $mu$ pairs for a range of network parameters and network types (fig.S.6). The correlation coefficients between mortality rates from ages $t0$  to $t0+5$ and $t_1$ to $t_1+5$ is $\rho\sim0.8, 0.98$ and $0.99$ for $\{t_0,t_1\}=\{0,20\},\{10,20\}$ and $\{15,20\}$ respectively. When determining these values the bin size was chosen as $5$ instead of $1$ in order reduce finite size effects (e.g. not many networks die \emph{exactly} on step-1).\\



\begin{thebibliography}{26}
\bibitem{foot2}
Analogous equations can be written for $\sigma^{(i)}$ for $i>1$; however these equations will depend on $P(i; x_1x_2\ldots x_i)$, the probability that a node with $i$ providers has one provider with $x_1$ providers, one provider with $x_2$ providers and so on. Fortunately we do not need these equations to obtain $\phi_c$ and $a$
\end{thebibliography}
\end{document}